# A Modified Model for Static Friction of a Soft and Hard Solid Interface

Arun K. Singh[1] and Vinay A. Juvekar[2]

[1]Department of Mechanical Engineering, Visvesvaraya National Institute of Technology, Nagpur-440010, India

[2]Department of Chemical Engineering, Indian Institute of Technology Bombay, Mumbai- 400076, India

E-mail[1]: aksinghb@gmail.com

**Abstract:**

We present a modified model based on shear rate and aging time dependent static friction between a soft solid such as gelatin hydrogel and a hard surface for instance glass surface. Earlier the model for static friction (Juvekar and Singh, 2016) considered only the bond rupture process as a result, the friction model over predicts the static friction in the experiment. The friction model now takes into account both formation and rupture of molecular chains at the sliding interface. It is also assumed that age of the newly formed bonds during the rupture process is the same as the aging time. As a result, the model predicts quite well the experimental data and thus highlighting the significance of bond formation in static friction. Moreover, it is also observed that residual stress has no effect on static strength.

## 1. Introduction

Static friction of a sliding surface is defined as the minimum shear force required to onset sliding of the block. Unlike dynamic friction, it is a history dependent property which depends on aging time as well as shear velocity (Person, 2000). This may be one of the reasons why prediction of static friction is a challenging problem in tribology. Present study involves static friction, where a soft block is allowed to age on a hard surface and then it sheared until that begins to slide. We determine the stress at the onset of sliding of the block using population balance equation (Singh and Juvekar, 2011). Earlier Juvekar and Singh (2016) studied this problem without considering bond formation process during the pulling of the soft block. But this proposed model over predicts the peak value of static friction.

Consider the case where the soft solid block is allowed to age on the hard surface for a time interval $t_w$, called the waiting time (aging time) or hold time, and then its upper face is pulled at a constant velocity $V_0$, in the direction parallel to the base. During the waiting time, polymer chains adsorb on the hard surface. The number of bonds, between the polymer chains and the hard surface, increases with increase in the waiting time. During the pulling stage, the block undergoes shear deformation. The resulting shear stress causes stretching of the chains, which are bonded to the hard surface. The force generated due to stretching of the chains exactly balances the pulling force. This stress also causes creep at the interface. The creep velocity, $V_c$, progressively increases with time due to the increase in shear stress as well as decrease in the number of the live bonds due to breakage. As long as $V_c < V_0$, the block continues to deform and the stress increases with time. A peak stress is reached when $V_c = V_0$. At this point, the stress is momentarily independent of time. We call this point, the point of onset of sliding, although the actual sliding of the block has begun much earlier. The peak stress is generally characterized as the stress of static friction (Baumberger et al., 2003). Beyond the peak, $V_c > V_0$, and both the deformation of the block and the consequent stress decrease.

We model this process as follows. During the waiting period or hold period, bonds are only formed. They neither age nor are they broken. Hence all bonds have zero age, and the population balance equation simplifies to

$$\frac{dN(t)}{dt} = \frac{1}{\tau}[N_0 - N(t)] \qquad (1)$$

Here, $N(t)$ is the total number of live bonds and $N_0$, the total number of available bonding sites. This equation yields the following solution subject to the initial condition, $N(t) = 0$ at $t = 0$.

$$N(t) = N_0\left[1 - exp\left(-\frac{t}{\tau}\right)\right] \qquad (2)$$

The number $N_i$ of bonds, formed at the end of the waiting time $t_w$ is

$$N_i = N_0\left[1 - exp\left(-\frac{t_w}{\tau}\right)\right] \qquad (3)$$

During the pulling process, the block is subjected to shear deformation. We analyze the dynamics of bonds during the time interval from the beginning of pulling to the time when the friction stress reaches the peak. We denote this time interval by $t_p$. The case of practical importance is the one for which $t_p << t_w$. Presently, we only analyze this case.

We earlier assumed that only those bonds, which are formed during the waiting period, bear the stress during the pulling period (Juvekar and Singh, 2016). We now take into account the contributions from the bonds which are newly formed during the pulling period. The resulting bonds formed are, therefore, weaker than those formed during the waiting time when the block was stationary. As a result, we can take into account the birth-term in Eq 1 assuming that newly formed bonds have almost the same strength. Also, since all bonds formed during the aging period have the same age, we can write $t_a = t$ for all bonds and simplify PBE (Eq 8) to the following form in terms of per chain force $f(t)$, adhesion constant $u$, time constant $\tau$, total number of chains $N_0$, number of chains adhered with the substrate $N$, using the Eyring's equation of rate reaction in view of bond formation as well as rupture at the sliding interface

$$\frac{dN(t)}{dt} = \frac{(N_0 - N(t))}{\tau} - N(t)\frac{u}{\tau}e^{\lambda f(t)/kT} \qquad (4)$$

$N(t)$ now represents the total number of live bonds at any time $t$ during the pulling phase. Negative sign on the right hand side of Eq 4 shows that $N(t)$ continuously decreases with time. We assume validity of the Hooke's law, and write

$$\frac{df(t)}{dt} = MV_c(t) \qquad (5)$$

The total force per unit area that is, frictional stress $\sigma(t)$ is given by

$$\sigma(t) = N(t)f(t) \qquad (6)$$

The stress $\sigma(t)$ also equal the force generated in the soft block having stiffness $K_b$ due to its shear deformation, and is obtained by the following equation

$$\frac{d\sigma(t)}{dt} = K_b \frac{d\Delta L(t)}{dt} = K_b(V_0 - V_c(t)) \qquad (7)$$

Differentiating Eq 6 with respect to time and simplifying the resulting equation using Eq 4 and 5, we obtain

$$\frac{d\sigma(t)}{dt} = \frac{\sigma(t)}{N(t)}\left\{\frac{(N_0 - N(t))}{\tau} - \frac{N(t)u}{\tau}\exp\left[\frac{\sigma(t)\lambda}{N(t)kT}\right]\right\} + MN(t)V_c(t) \qquad (8)$$

Eliminating $d\sigma(t)/dt$ between Eqs 7 and 8, we obtain the following expression of the creep velocity

$$V_c(t) = \frac{K_b V_0 - \dfrac{f(t)(N_0 - N(t))}{N(t)\,\tau} + f(t)\left(\dfrac{u}{\tau}\right)e^{\lambda f(t)/kT}}{MN(t) + K_b} \tag{9}$$

substitution of this expression into Eq 7 yields

$$\frac{d\sigma(t)}{dt} = \frac{\sigma(t)}{N}\left\{\frac{[N_0 - N(t)]}{\tau} - \frac{N(t)u}{\tau}\exp\left[\frac{\sigma(t)\lambda}{N(t)kT}\right]\right\} + MN(t)V_c(t) \tag{10}$$

We eliminate $f(t)$ from Eqs 4 and 10 using Eq 6, and modify them respectively to the following forms

$$\frac{dN(t)}{dt} = \frac{(N_0 - N(t))}{\tau} - N(t)\frac{u}{\tau}e^{\lambda\sigma(t)/N(t)kT} \tag{11}$$

$$\frac{d\sigma(t)}{d\hat{t}} = K_b N(t)\left(\frac{\dfrac{(N_0 - N(t))}{\tau} - \dfrac{u}{\tau N(t)}\sigma(t)\exp\left(\dfrac{\lambda\sigma(t)}{N(t)kT}\right)}{N(t)M + K_b}\right) \tag{12}$$

The ordinary differential equations (Eqs 11 and 12) now contain only two dependent variables $\sigma(t)$ and $N(t)$. They can be simultaneously solved using the following initial conditions

$$N(0) = N_i \quad \text{and} \quad \sigma(0) = 0 \tag{13}$$

where, $N_i$ is the number of bonds formed at the end of the aging period, and is given by Eq3.

We now use the dimensionless transformations given by Eq 3, 7 and 6 to render Eqs 11 and 12 dimensionless. Moreover, we also write

$$\hat{N} = \frac{N}{N_0}, \quad X = \frac{N_i - N(t)}{N_i} \text{ or } N(t) = (1-X)\hat{N}_i \text{ and } r_s = \frac{N_0 M}{K_b} \tag{14}$$

In the above equation, $X$ is the fraction of the initial bonds which are broken at time $t$ and $r_s$ is the ratio of the total stiffness of the chains to the stiffness of the bulk material. Using these transformations, Eqs 11 and 12 can be modified to

$$\frac{dX}{d\hat{t}} = -r_s \hat{N}_i^{-1}\left[1-(1-X)\hat{N}_i\right] + r_s u(1-X)\exp\left[\frac{\hat{\sigma}(\hat{t})}{(1-X)\hat{N}_i}\right]$$

$$\frac{d\hat{\sigma}(t)}{d\hat{t}} = \frac{\hat{V}_0(1-X)\hat{N}_i + \hat{\sigma}(t)\left[(1-X)^{-1}\hat{N}_i^{-1}-1\right] - \hat{\sigma}(t)u\exp\left[\frac{\hat{\sigma}(t)}{(1-X)\hat{N}_i}\right]}{(1-X)\hat{N}_i + 1/r_s} \quad (15)$$

The initial conditions given by Eq 13 transform to

$$X(0)=0 \ and \ \hat{\sigma}(0)=0 \quad (16)$$

The velocity of creep can be written in dimensionless form as

$$\hat{V}_c(t) = \frac{\hat{V}_0 - \left[\frac{1}{(1-X)\hat{N}_i} - 1\right] + u r_s \hat{\sigma}(t)\exp\left(\frac{\hat{\sigma}(\hat{t})}{(1-X)\hat{N}_i}\right)}{r_s \hat{N}_i(1-X)+1} \quad (17)$$

The system of equations in Eq15 may also be written as

$$\frac{dX}{d(u\hat{t})} = -r_s \hat{N}_i^{-1} u^{-1}\left[1-(1-X)\hat{N}_i\right] + r_s(1-X)\exp\left[\frac{\hat{\sigma}(\hat{t})}{(1-X)\hat{N}_i}\right]$$

$$\frac{d\hat{\sigma}(t)}{d(u\hat{t})} = \frac{\left(\frac{\hat{V}_0}{u}\right)(1-X)\hat{N}_i + \hat{\sigma}(t)\left[(1-X)^{-1}\hat{N}_i^{-1}-1\right] - \hat{\sigma}(t)u\exp\left[\frac{\hat{\sigma}(t)}{(1-X)\hat{N}_i}\right]}{(1-X)\hat{N}_i + 1/r_s} \quad (18)$$

**2. Experimental validation and development of the scaling laws:**

We have also validated the friction model (Eqs 19&20) on gelatin gel/ glass interface. The value of friction parameters are determined after the best curve fit of the friction model (Eq 19&20) with static friction vs. time for $V_0 = 1 mms^{-1}$ in Fig.1. The value of the friction parameter is $\sigma^* = 3.6 kPa$, $V_* = 10^{-3}$ $mms^{-1}$, $u_0 = 1$, $N_i = 1$, $r_s = 50$ and gel stiffness $K_b = 600$ $kPam^{-1}$. Fig.1 also shows that peak of static friction is quite closure to the experimental value for $V_0 = 2 mms^{-1}$ to 5 $mms^{-1}$. It is also important to point that residual stress is not considered in the present analysis. As it is believed that since magnitude of residual stress is of the same magnitude, so is not important in determining the relative magnitude of static friction. However it may be taken into account during the absolute value of static friction.

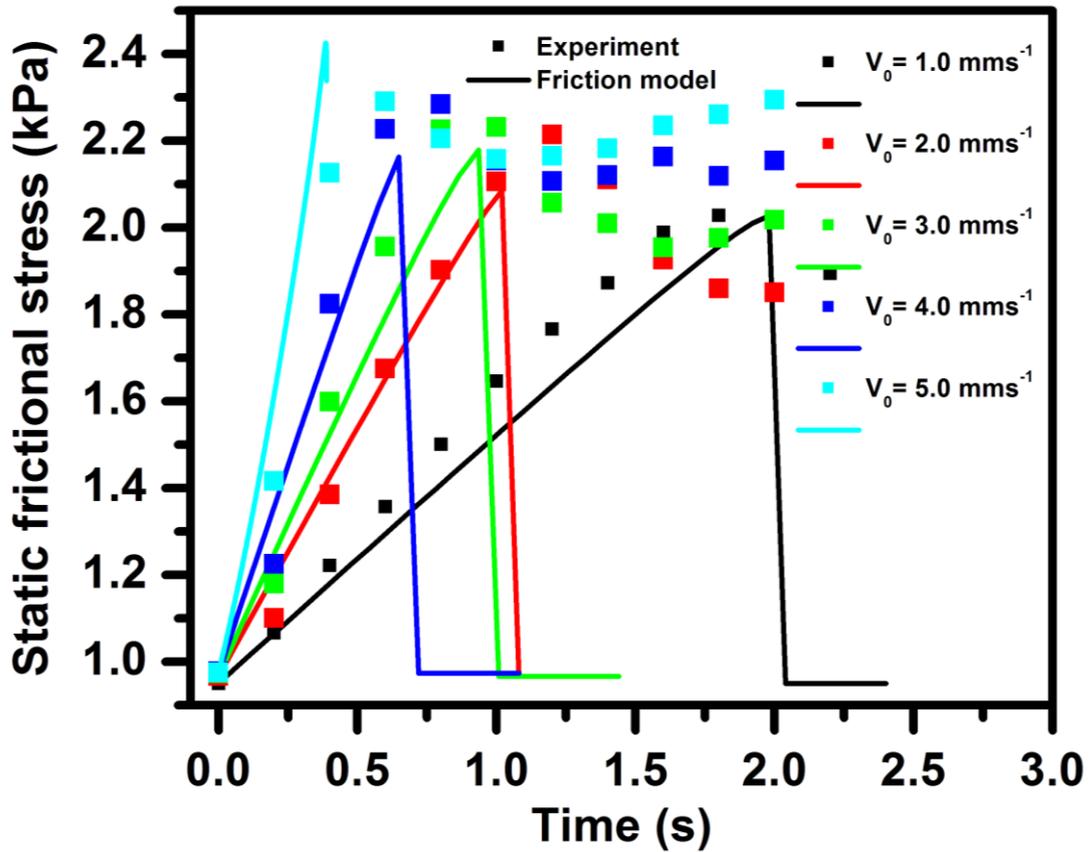

Fig.1 A match between the experimental static frictional stress (discrete points) and the static friction model (continuous line) of gelatin hydrogel c=10% (wt./vol.) at different pulling velocity $V_0$= 1,2,3,4 and 5mms$^{-1}$.

Further validation of the friction model in view of more experimental data would be interesting. It would also be interesting to validate the experimental data concerning the effect of aging time/hold time on static friction (Li et al., 2011).

**5. Conclusion:** An important observation is that residual stress has no effect on static strength. While the modified friction model predicts well the experimental data for static friction, life of the newly formed bond was considered the same as the life of all bonds before the experiments. Moreover, a future work may be perused considering the role of FENE chain on static strength of soft polymer surfaces.

# References

(1) Baumberger, T.; Caroli, C.; Ronsin, O. Self-healing pulses and friction of gelatin gels. *Eur. Phys. J. E* **2003**, 11, 85-93.

(2) Juvekar, V.A. and Singh, A.K. Rate and aging time dependent static friction of a soft and hard solid interface. **2016**, *arXiv preprint arXiv:1602.00973*.

(3) Li, Q;Tullis, T.E.; Goldsby, D.;Carpick, R.W. Frictional ageing from interfacial bonding and the origins of rate and state friction. *Nature* **2011**, 480, 233-236.

(4) Persson, B.N.J. Sliding friction: physical principles and applications; Springer-Verlag: Berlin and Heidelberg, **2000**.

(5) Singh, A.K.; Juvekar, V.A. Steady dynamic friction at elastomer-hard solid interface: A model based on population balance of bonds. *Soft Matter* **2011**,7, 1060101-1060111.